\documentstyle[12pt]{article}
\begin{document}
\baselineskip=24pt
\title{
\vspace{-3.0cm}   
\begin{flushright}   
{\normalsize UTHEP-280}\\ 
\vspace{-0.3cm} 
{\normalsize July 1994 }\\
\end{flushright}
\vspace*{2.0cm}
{\Large Manifestation of Sea Quark Effects in the Strong Coupling Constant in
Lattice QCD\vspace*{0.5cm}} } 
\author{S. Aoki$^1$, M. Fukugita$^{2\dagger}$,  S. Hashimoto$^3$, N.
Ishizuka$^1$,\\ 
        H. Mino$^4$, M. Okawa$^5$,  T. Onogi$^{3\dagger\dagger}$ and A.
Ukawa$^1$\\ \\        
{\it $ ^1$Institute of Physics, University of Tsukuba,}\\ 
{\it      Tsukuba, Ibaraki-305, Japan} \\
{\it $ ^2$Yukawa Institute for Theoretical Physics, Kyoto University,}\\
{\it     	Kyoto 606, Japan }\\
{\it $ ^3$Department of Physics, Hiroshima University,}\\ 
{\it      Higashi-Hiroshima 724, Japan}\\ 
{\it $ ^4$Faculty of Engineering, Yamanashi University,}\\
{\it      Kofu 400, Japan} \\ 
{\it $^5$National Laboratory for High energy Physics(KEK),}\\
{\it     Ibaraki 305, Japan}
}
\date{}
\maketitle

\begin{abstract}
\baselineskip=24pt
We demonstrate that sea quark effects of a magnitude expected from
renormalization group considerations are clearly visible in the strong
coupling constant measured in current full QCD simulations.   Building on this
result an estimate of $\alpha_{\overline{MS}}^{(5)}(M_Z)$ is made employing the
charmonium $1S-1P$ mass splitting calculated on full QCD configurations
generated with two flavors of dynamical Kogut-Susskind quarks to fix the scale.
\end{abstract}
\vfill

\hrule
\vskip 0.20cm
\baselineskip=18pt
\noindent ${}^\dagger$ Also at {\it Institute for
Advanced Study, Princeton, NJ 08540, U. S. A.}\\
${}^{\dagger\dagger}$ Presently at {\it Theory Group, Fermi National Accelerator
Laboratory,\\\hspace*{5mm} Batavia, IL 60510, U. S. A.}\\

\baselineskip=24pt

\newpage

A distinctive feature of QCD that differentiates it from
phenomenological quark models of hadrons is the existence of sea quarks
built into the theory.  Nonetheless, finding a physical manifestation of sea
quark effects has been an elusive subject in full lattice QCD simulations.  In
hadron mass spectrum calculations, for example, full QCD results for flavor
non-singlet hadron masses agree with those of quenched QCD  within statistical
errors of 5--10\% if the bare coupling constant for the latter is shifted by an
appropriate amount\cite{fou}.  A similar situation holds for the critical
coupling of the chiral transition at finite temperatures in full QCD; its
value, though largely dependent on the sea quark mass, is reproduced quite well
from that of the pure gauge theory by correcting for quark one-loop vacuum
polarization effects\cite{hasenfratzdegrand}.   

A possible interpretation of the matching of
full and quenched QCD by a shift of the bare coupling is that the shift
represents an adjustment of the renormalized coupling constant at the low energy
scale that dominates the behavior of quantities being simulated\cite{EK,NRQCD}.
If this interpretation is valid, one expects that sea quark effects will become
manifest in the renormalized coupling constant estimated for a scale
sufficiently large compared to the dominant scale, since the full QCD coupling
constant decreases more slowly than that of the pure gauge theory.  In this
article we present evidence that this in fact is the case: we find that the full
QCD coupling constant extracted from two-flavor full QCD simulations is
consistently larger than that of quenched QCD at large momenta ranging over
$\mu\approx 3-7$GeV when the scale is determined from the $\rho$ meson mass;
the  difference in the magnitude of the full and quenched coupling
constants is consistent with the picture that the two
couplings merge when evolved down to the low energy scale $\mu<1$GeV via the
two-loop renormalization group. We also estimate the physical strong coupling
constant for five flavors $\alpha_{\overline{MS}}^{(5)}$ at the $M_Z$ scale
following the work of Ref.~\cite{EK,NRQCD} employing the $1S-1P$ mass splitting
of charmonium states estimated for two-flavor full QCD.

Calculation of the renormalized value of the coupling for a bare value
$\alpha_0$ taken in a simulation is facilitated by the recent study\cite{LM}
which has shown  that lattice perturbative series is well convergent after
lattice gluon tadpole effects are properly taken into account.  A proposal for
including  tadpole effects in the relation between the bare and renormalized
coupling is given by\cite{EK}
\begin{equation}
\alpha^{(N_f)}_{\overline{MS}}(\pi/a)^{-1} 
     = P\alpha_0^{-1} + c_g + N_fc_f+O(\alpha_0^2),
\label{eq:MSbar}
\end{equation}
where $P$ is the plaquette expectation value, $c_g=0.30928$ is the gluon
one-loop contribution\cite{HHDG}, and the last term represents the
contribution of $N_f$ flavors of quarks with $c_f=-0.08848$ for the
Kogut-Susskind quark action\cite{STW} and $c_f=-0.03491$ for the Wilson
action\cite{KNSweisz}.    Alternatively one may use the $\alpha_V$
coupling defined from the static $\bar q q$ potential, which can be
estimated from $P$ via\cite{LM}  
\begin{equation}
-\log P=\frac{4\pi}{3}\alpha^{(N_f)}_V(3.41/a)\left(
1+(d_g+N_fd_f)\alpha^{(N_f)}_V+O(\alpha_V^2)\right)
\label{eq:V}
\end{equation}
with $d_g=-1.1855$ and $d_f=-0.0703$ for the Kogut-Susskind quark action and
$d_f=-0.0249$ for the Wilson action. The relation between the two couplings are
given by\cite{LM,bill}
\begin{equation}
\alpha^{(N_f)}_{\overline{MS}}(\mu)^{-1}=\alpha^{(N_f)}_V(\mu)^{-1}
+c_{\overline{MS}}^V+O(\alpha^{(N_f)}_V(\mu)^2),\qquad 
c_{\overline{MS}}^V=\frac{1}{36\pi}\left(93-10N_f\right).
\label{eq:VMS}
\end{equation}
Equivalently the $\Lambda$ parameters are related by 
\begin{equation}
\Lambda_{\overline{MS}}^{(N_f)}=
\exp(-\frac{c_{\overline{MS}}^V}{8\pi b_0})\Lambda_V^{(N_f)} 
\end{equation}
with $b_0=(11-2N_f/3)/(4\pi)^2$.

The renormalized coupling constant in the $\overline{MS}$ scheme extracted
from (\ref{eq:MSbar}) for quenched and two-flavor full QCD are compared 
in Fig.~\ref{fig:fig1}  for Kogut-Susskind and Wilson quark actions as a
function of scale $\mu=\pi/a$ determined from the $\rho$ meson mass. In full QCD
the plaquette data are extrapolated linearly in the sea quark mass to $m_q=0$.
In quenched QCD we made a ninth order polynomial fit in $\beta$  of plaquette
values published in the literature\cite{quenchedplaquette} in order to calculate
the values at $\beta$ where data are not available.  The
trend is apparent in Fig.~\ref{fig:fig1} that the full QCD coupling constant is
systematically larger than that of the pure gauge theory when compared at the
same scale $\mu$.   

The solid lines in Fig.~\ref{fig:fig1} illustrates the two-loop renormalization
group evolution of the coupling constant.  Deviation of 
$\alpha^{(N_f)}_{\overline{MS}}(\pi/a)$ from the solid lines toward smaller
values of cutoff is in part ascribed to scaling violation effects due to a finite
lattice spacing and in part to uncertainties of $O(\alpha^2)$ in the relation 
(\ref{eq:MSbar}).  One can estimate the magnitude of the latter through a
comparison of the coupling constant extracted  from 
(\ref{eq:MSbar}) and (\ref{eq:V}).  This analysis shows that the latter estimate
yields values for $\alpha^{(N_f)}_{\overline{MS}}(\pi/a)$ larger by about
3--5\% at $\mu\approx 7$GeV, and by about 5--10\% at $\mu\approx 3$GeV. This
is taken as uncertainties of our analyses. 

In Fig.~\ref{fig:fig2} we compare the two-loop renormalization group evolution
of the full and quenched coupling constants toward small momenta
$\mu<0.5-1$GeV.  The upper and lower edges of the bands in this figure
correspond to  $\alpha^{(N_f)}_{\overline{MS}}(\pi/a)$ estimated from the
relation (\ref{eq:MSbar}) and that
from (\ref{eq:V})  including scale errors in order to take
into consideration the two-loop uncertainty.  For full QCD we
employ the data taken at the highest $\beta$ for the starting value, and for
quenched QCD the one carried out at a value of $\beta$ with a nearby value of
$\mu=\pi/a$.   We observe that the evolution of the two coupling constants
overlaps below $\mu\approx 0.4$GeV, which is the dominant scale relevant for the
$\rho$ meson that is employed for fixing the scale.  

The results described above are fully consistent with the view that matching full
and quenched results means adjusting the coupling constant at the relevant low
energy scale, and that the full QCD couplings estimated for larger momenta 
should exhibit a slower decrease than the pure gauge
coupling with a rate dictated by the renormalization group $\beta$ function.

Let us note that these findings provide support for the procedure
of Refs.~\cite{EK,NRQCD} for estimating the physical strong coupling constant
from values measured in simulations with an incomplete spectrum
of sea quarks.    Namely one initially evolves the measured value at the cutoff
scale down to the low energy scale $\mu_0$ typical of the
simulated hadron system using the flavor number of the simulation.  The physical
coupling constant $\alpha_s$ at that scale is equated to the evolved value, and
$\alpha_s$ for larger scale is calculated through renormalization group
incorporating the full spectrum of quarks active at each scale.   A necessary
condition for applying this procedure is that the scale $\mu_0$ is not too small
in order not to spoil the two-loop approximation to the $\beta$ function
that breaks down for small momenta.  The authors of Ref.~\cite{EK} proposed to
use the $1S-1P$ charmonium mass splitting for which potential models suggest 
$\mu_0\approx 0.4-0.75$GeV.  The advantage is that heavy
quark propagators are easy to calculate and that the $1S-1P$ mass splitting is
empirically insensitive to the quark mass, rendering its fine tuning unnecessary.

We have carried out an analysis along this line employing  the full QCD
configurations on a $20^4$ for two flavors of
dynamical Kogut-Susskind quarks  at $\beta=5.7$ with $m_qa=0.01$\cite{fmou}.  
For charmonium spectrum measurement we used the Wilson quark action for valence
quarks\cite{kscharm}, employing Gaussian smeared sources
$\sum_{x,y}\bar\psi_x\Gamma_1\psi_yf(x)f(y)$ and local sinks
$\bar\psi_x\Gamma_2\psi_x$ where $\Gamma_1=\Gamma_2=\gamma^i$,
$\gamma^5$, $\sigma^{jk}$ for $J/\psi$, $\eta_c$, $h_c(op.~1)$ with  $f(x)=\exp
(-\vert x\vert^2/4)$, and also $\Gamma_1=\gamma^5$, $\Gamma_2=\sigma^{jk}$
 with $f(x)=\sin (2\pi x^i/L)f_s(x)$ for $h_c(op.~2)$.  We analyzed 72
configurations at $K=0.130$ and 75 configurations at $K=0.135$ with the lattice
size periodically doubled in the temporal direction. 

In Table~\ref{tabletwo} we list our result for the
charmonium spectrum and the corresponding scale $\pi/a$ extracted from
the experimental value of the $1P-1S$  mass splitting 
457.8(5)MeV\cite{onepones} as input. The splitting is almost independent of the
hopping parameter $K$ which controls the charm quark mass, though it slightly
depends on the choice of the operator for $h_c$ ($op.~1$ or $op.~2$).
The splitting yields a value $\pi/a\approx 7$GeV, which is
consistent with $\pi/a=7.01(28)$ \cite{fmou} estimated from the $\rho$ meson
mass.  

Our results for the physical strong coupling constant and the $\Lambda$
parameter obtained with the scale listed in Table~\ref{tabletwo} are summarized
in Table~\ref{tablefour}.  In  Table~\ref{tablefour}(a) the starting value is 
$\alpha_{\overline{MS}}^{(2)}(\pi/a)=0.142$ estimated from (\ref{eq:MSbar}),
while in  Table~\ref{tablefour}(b) we use $\alpha_V^{(2)}(3.41/a)=0.169$
obtained from (\ref{eq:V}).  For both cases the actual evolution is made in terms
of the $\alpha_V$ coupling since it is directly related to the heavy quark
potential relevant for charmonium.
We use  $\mu_0=0.4$ GeV and 0.75 GeV for the matching scale and take an average
of the two choices for the central value for the strong coupling constant.
The errors are estimated by allowing $\mu_0$ to vary over the range
$\mu_0=0.4-0.75$GeV and the scale $\pi/a$ within the quoted error.  

Our results are consistent with the previous lattice estimates carried out in
quenched QCD\cite{EK,NRQCD}. Compared to the world average of
phenomenological determinations
$\alpha_{\overline{MS}}^{(5)}(M_Z)=0.118(7)$\cite{bethke}, the results
are somewhat small especially for those estimated from
the relation (\ref{eq:MSbar}).  We do
not view the difference to be alarming at
this stage since there exists a 5\%
uncertainty in our value of $\alpha_{\overline{MS}}^{(5)}(M_Z)$ due to that of
the input value at the cutoff scale, and an additional 5\% that results
from the matching procedure, as well as errors in the experimental value.

To summarize, our analyses have shown that sea quark effects of a magnitude
expected from renormalization group considerations are visible in the strong
coupling constant measured in current full QCD simulations incorporating up and
down quarks.  This indicates a promising prospect for a realistic determination 
of the strong coupling constant including the full spectrum of sea quarks
since incorporating  heavy quarks such as  strange and charm is not difficult 
from the view of the necessary computer power compared to that for light quarks.

\newpage
\section*{Acknowledgements}

Numerical calculations for the present work have been
carried out on  HITAC S820/80 at KEK. 
One of us (T.O.) thanks A. El-Khadra,  A. Kronfeld and P. Mackenzie  
for informative discussions.
This work is supported in part by the Grants-in-Aid of 
the Ministry of Education(Nos. 03640270, 04NP0601, 05640325, 05640363).

\newpage
                                    
\begin{table}[h]
\begin{center}
\begin{tabular}{lllllll}
\hline\\[-3mm] 
$K$ & $m_{\eta_c}a$ & $m_{J/\psi}a$ & &$ m_{h_c}a$ & $\Delta m_{1P-1S}a $ 
&  $\pi/a$(GeV)  \\[0.2cm]
\hline\\[-3mm]
0.130 & 1.474(16) & 1.466(17) &op.~1:& 1.692(53) & 0.221(49) & 6.5(1.5)\\[0.2cm]
      &           &           &op.~2:& 1.672(32) & 0.200(32) & 7.2(1.2)\\[0.2cm]
0.135 & 1.285(16) & 1.275(15) &op.~1:& 1.504(53) & 0.221(48) & 6.5(1.5)\\[0.2cm]
      &           &           &op.~2:& 1.487(32) & 0.205(35) & 7.0(1.2)\\[0.2cm]
\hline
\end{tabular}
\caption{Charmonium spectrum for Wilson  valence quarks in full QCD
at $\beta=5.7$ with two flavors of dynamical Kogut-Susskind quarks with
$m_qa=0.01$. } \label{tabletwo}    \end{center}
\end{table}

\begin{table}[h]
\begin{center}
\begin{tabular}{lllll}
\hline\\[-3mm]
  $K$ & $\alpha_{\overline{MS}}^{(4)}(\mbox{5GeV})$
& $\Lambda_{\overline{MS}}^{(4)}$ & $\alpha_{\overline{MS}}^{(5)}(m_Z)$
& $\Lambda_{\overline{MS}}^{(5)}$
\\[0.2cm]
\hline\\[-3mm]
\multicolumn{5}{l}{(a) estimate from
$\alpha_{\overline{MS}}^{(2)}(\pi/a)$}\\[3mm] 
 0.130($op.~1$) & $0.168^{+0.015}_{-0.010}$ &
		$141_{-34}^{+57}$ MeV & $0.104^{+0.005}_{-0.004}$ &
 $92_{-24}^{+43}$ MeV\\[0.2cm]
 0.135($op.~1$) & $0.168_{-0.010}^{+0.015} $& 
	$ 141_{-33}^{+55}$ &$0.104_{-0.004}^{+0.005}$ &
$92_{-24}^{+42}$ MeV \\[0.2cm]
 0.130($op.~2$) & $0.173_{-0.009}^{+0.012}$ &
	  $159_{-31}^{+44}$ & $0.106_{-0.003}^{+0.004}$ &
$105_{-22}^{+34}$ MeV\\[0.2cm]
 0.135($op.~2$) & $0.172_{-0.009}^{+0.012}$ & 
	$154_{-31}^{+46}$ & $0.106_{-0.003}^{+0.004}$ &
$102_{-23}^{+35}$ MeV\\[0.2cm]
\hline\\[-3mm]
\multicolumn{5}{l}{(b) estimate from $\alpha_V^{(2)}(3.41/a)$}\\[3mm] 
 0.130($op.~1$) & $0.186^{+0.017}_{-0.013}$ &
		$206_{-50}^{+74}$ MeV & $0.110^{+0.006}_{-0.005}$ &
$142_{-36}^{+57}$ MeV\\[0.2cm]
 0.135($op.~1$) & $0.186_{-0.013}^{+0.016} $& 
	$ 206_{-49}^{+72}$ &$0.110_{-0.005}^{+0.006}$ &
$142_{-38}^{+56}$ MeV\\[0.2cm]
 0.130($op.~2$) & $0.192_{-0.011}^{+0.013}$ &
	  $231_{-44}^{+57}$ & $0.112_{-0.004}^{+0.004}$ &
$161_{-34}^{+45}$ MeV \\[0.2cm]
 0.135($op.~2$) & $0.190_{-0.011}^{+0.014}$ & 
	$224_{-44}^{+60}$ & $0.112_{-0.004}^{+0.004}$ &
$156_{-35}^{+47}$ MeV\\[0.2cm]
\hline 
\end{tabular}
\caption{Strong coupling constant
and $\Lambda$ parameter in the ${\overline{MS}}$ scheme 
according to the Particle Data
Group definition 
calculated with the $1S-1P$ charmonium mass splitting with Wilson valence
quarks for fixing the scale.}  \label{tablefour}   \end{center} \end{table}

\newpage

\section*{Figure captions}

\begin{figure}[h]
\caption{Comparison of $\alpha^{(N_f)}_{\overline{MS}}(\pi/a)$ estimated via 
(\protect\ref{eq:MSbar})  for
two-flavor full QCD (filled symbols) and quenched QCD (open symbols)  with the
scale fixed by the $\rho$ meson mass.} \label{fig:fig1} \end{figure}

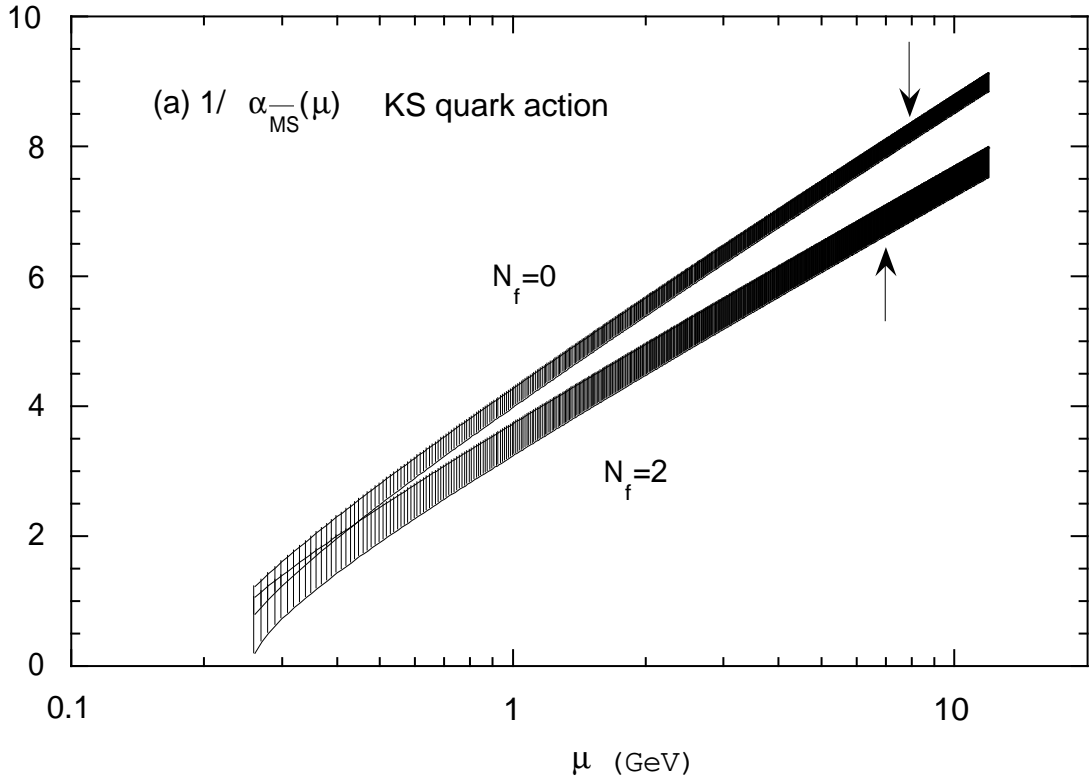
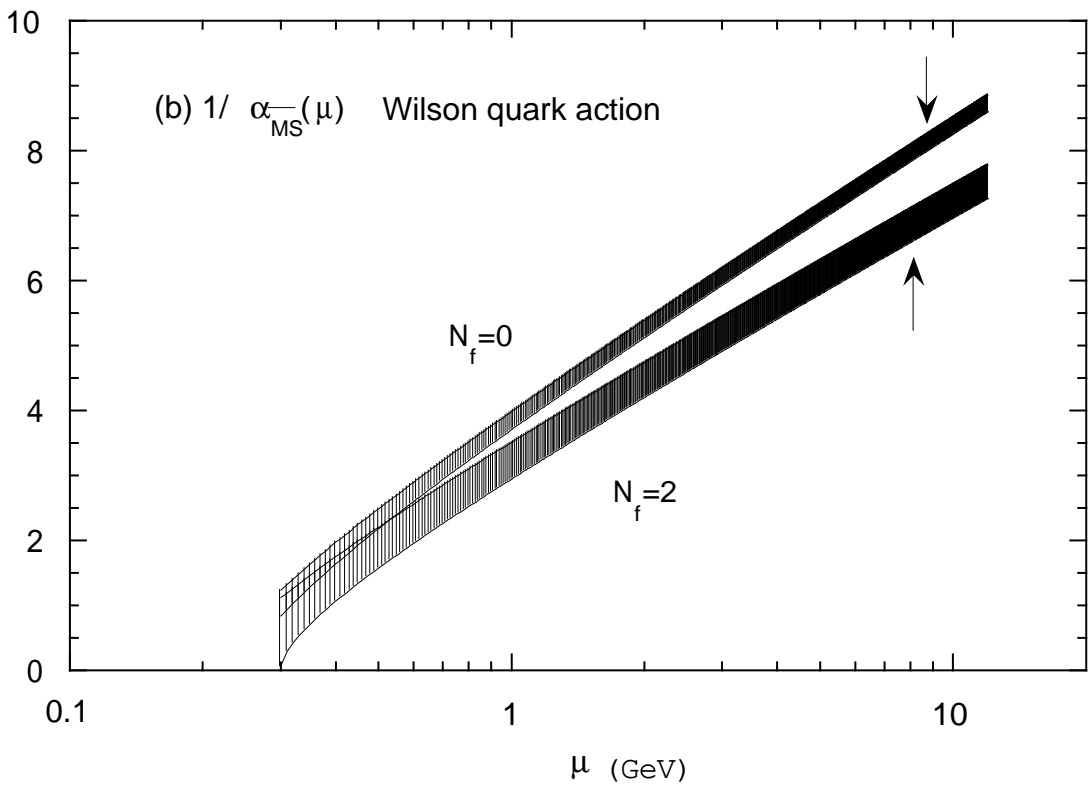
\begin{figure}[h]
\caption{Evolution of $\alpha^{(N_f)}_{\overline{MS}}(\mu)$ for quenched and two-flavor
QCD. Bands correspond to uncertainties in the estimate of
$\alpha^{(N_f)}_{\overline{MS}}(\mu)$. Arrows indicate the starting value
taken from (a) Fukugita {\it et al.} $(N_f=2, \beta=5.7)$\protect\cite{ksnfull}
and   Sharpe {\it et al.}$(N_f=0, \beta=6.2)$\protect\cite{ksquenched}, and (b)
Gupta {\it et al.}$(N_f=2, \beta=5.6)$\protect\cite{wilsonfull} and Butler {\it
et al.}$(N_f=0, \beta=6.17)$\protect\cite{wilsonquenched}. }   
\label{fig:fig2} \end{figure} \vfill


\begin{thebibliography}{99}
\bibitem{kscharm}A similar analysis employing the Kogut-Susskind quark action
for valence quarks are currently in progress, and will be published elsewhere.
\end{thebibliography}
\end{document}